\documentclass[superscriptaddress,prd,reprint,floatfix,nofootinbib,longbibliography]{revtex4-1}

\usepackage[T1]{fontenc}
\usepackage[utf8]{inputenc}
\usepackage{cancel}

\usepackage{amssymb,amsmath}
\usepackage{graphicx}
\usepackage{subfigure}
\usepackage{color}
\usepackage[dvipsnames]{xcolor}
\usepackage[colorlinks]{hyperref}
\hypersetup{
	colorlinks=true, 
	linktoc=all,     
	linkcolor=blue,  
	citecolor=magenta,
	urlcolor=blue	
}
\usepackage{dcolumn}
\usepackage{bm}
\usepackage{interval}
\usepackage{rotating}
\usepackage{multirow}
\usepackage[separate-uncertainty = true,multi-part-units=single,multi-part-units=brackets]{siunitx}
\usepackage{url}
\usepackage[ruled,vlined]{algorithm2e}
\usepackage{cancel}
\usepackage{tikz}
\usetikzlibrary{arrows,decorations.markings}
\usepackage{tikz-3dplot}
\usetikzlibrary{math} 
\usetikzlibrary{calc}
\usepgflibrary{arrows}
\tikzset{
	partial ellipse/.style args={#1:#2:#3}{
		insert path={+ (#1:#3) arc (#1:#2:#3)}
	}
}

\def\centerarc[#1](#2)(#3:#4:#5)
{ \draw[#1] ($(#2)+({#5*cos(#3)},{#5*sin(#3)})$) arc (#3:#4:#5); }

\begin{document} 

\date {\today}

\title{Quantum mechanics of radiofrequency-driven coherent beam oscillations \\ in storage rings} 

\author{J. Slim}
\thanks{Now at: GSI Helmholtzzentrum für Schwerionenforschung, 64291 Darmstadt, Germany}
\affiliation{III. Physikalisches Institut B, RWTH Aachen University, 52056 Aachen, Germany}

\author{N.N. Nikolaev}
\affiliation{L.D. Landau Institute for Theoretical Physics, 142432 Chernogolovka, Russia}
\affiliation{Moscow Institute for Physics and Technology, 141700 Dolgoprudny, Russia}

\author{F. Rathmann}
\affiliation{Institut f\"ur Kernphysik, Forschungszentrum J\"ulich, 52425 J\"ulich, Germany}

\author{A. Wirzba}
\affiliation{Institut f\"ur Kernphysik, Forschungszentrum J\"ulich,  52425 J\"ulich, Germany}
\affiliation{Institute for Advanced Simulation, Forschungszentrum J\"ulich, 52425 J\"ulich, Germany}




\begin{abstract}
Precision searches for the electric dipole moment of protons in storage ring experiments call for beam-position monitoring in the picometer range. We present the relevant  fully quantum mechanical derivation of 
radiofrequency-driven collective oscillations of a beam in a storage ring.  
\end{abstract}	
\maketitle

{\bf Introduction}	
One of grand challenges in particle physics is the search for physics beyond the Standard Model (BSM). At the forefront of the high precision frontier is the proposed search for the electric dipole moment (EDM) of protons stored in an  all-electric frozen-spin storage rings with a sensitivity of $d_p \sim \SI{e-29}{e.cm}$ that is some 15 orders of magnitude smaller than the magnetic dipole moment of the proton\,\cite{srEDM,CPEDM,OmarovHybrid}. The primary motivation is that the experimental observation of a permanent EDM of any subatomic particle implies the explicit violation of time reflection (T) and parity (P)  symmetries, and therefore, according to the CPT theorem, also involves the violation of CP in the flavor-preserving channel. The presence of the latter could shed light on the mystery of the anomalously large baryon asymmetry in the Universe,  which vastly exceeds the expectations within the Standard Models of particle physics and cosmology\,\cite{BernreutherBaryogenesis,BodekerBaryogenesis}.  

Proper control of systematic effects encountered in the search for such a minuscule EDM requires concurrent measurements of the spin rotations  of beams propagating in opposite directions in an all-electric ring. To achieve such an ambitious goal, one of the crucial tasks is to control the difference of the vertical positions of the two beams along the orbit in the machine with an accuracy of about $\SI{5}{\pico \meter}$\,\cite{CPEDM}. One may wonder whether such enormously demanding accuracy is not prohibited by the Heisenberg uncertainty principle. Moreover, this accuracy is also in the range of the vertical displacement of the beams due to Earth's  gravitational force. It should be noted that up to now  such tiny gravitational effects were never considered in the design and construction of accelerators, but gravity causes an observable background in the searches for the proton EDM\,\cite{SilenkoTeryaev2006,OrlovGravity,ObukhovSilenkoTeryaev,nikolaev2018gravity}. 
	
Recently, the first direct measurement of the amplitude of collective oscillations in the micrometer range of an intense beam of deuterons in a storage ring, excited by an intentionally  mismatched radiofrequency (RF) Wien filter (WF) \cite{Slim:2016pim} at the Cooler Synchrotron COSY of Forschungszentrum J\"ulich  has been reported \cite{Slim:2021wyk}. The achieved accuracy is about a factor of 10 larger than the amplitude of the zero-point betatron oscillation of \textit{single} particles. The latter is  $\approx \SI{41}{\nano \meter}$ (see\,\cite[Eq.\,(9)]{Slim:2021wyk}).
	
The approach to the quantum ground state and the possibility of detecting displacements of macroscopic bodies near the quantum limit are the subject of intense theoretical and experimental efforts\,\cite{Schreppler1486, Abbott:2009zz, Murch2008, Biercuk2010, Rugar2004}. A notable example is the detection of gravitational waves with interferometric detectors using kilogram-scale mirrors\,\cite{Abbott:2016izl}. The case of ultra-small coherent oscillations of particle bunches of rarefied gas confined in the focusing fields of storage rings is complementary to, and different from, the above examples. The possibility of detecting small amplitudes arises from the fact that the signal of {\em coherent} oscillations is generated by  $\approx \num{e9}$ particles contained in the bunch.
	
The subject of the present paper is the transition from the description of classical mechanics suitable for micron amplitudes in the COSY experiment\,\cite{Slim:2021wyk} to an obviously deep quantum regime of picometer amplitudes in the proposed ultimate proton EDM experiment. Our main conclusion is that the functional form of the coherent beam oscillation amplitude does note change during the transition from the classical to the quantum regime. It follows that considerations of the Heisenberg quantum uncertainty do not preclude sub-picometer accuracy in the determination of the coherent beam oscillation amplitude -- the challenge lies in the sensitivity of the beam position monitors (BPM)\,\cite{Kawall}. We also address the effects of intrabeam scattering and interactions with the residual gas on the coherent excitations of the beam. We illustrate the origin of the picometer domain by the example of the derivation of the vertical beam displacement by the Earth's gravity in terms of the vertical betatron frequency. 
	
{\bf Vertical splitting of counter-propagating beams due to Earth's gravity}
As an introduction into the subject, we explain how picometer-scale beam displacements emerge in the EDM experiment in storage rings. The attraction of the Earth on the beam particles  can not be switched off and  must be compensated for by the focusing electromagnetic fields: either radial magnetic fields in the case of focusing by magnetic quadrupoles \cite{SilenkoTeryaev2006,OmarovHybrid}, or vertical electric fields in the case of electric focusing\,\cite{ObukhovSilenkoTeryaev,OrlovGravity,srEDM, CPEDM}. The force of the gravitational attraction is given by
\begin{equation}
	\bm{F}_g = {\frac {2\gamma^2 - 1}{\gamma}}\,m\bm{g}_\oplus\, ,\label{Fg}
\end{equation}
where $\bm{g}_\oplus$ is the acceleration of free fall at the Earth's surface, $\gamma=1/\sqrt{1-(v/c)^2}$ is the Lorentz factor of the stored particles in the ring having velocity $v$, and $m$ is the rest mass of each particle. The spring constant $\langle k\rangle$ of  a confining oscillator potential can be related to the angular velocity $\omega_y$ of the vertical betatron oscillation as follows, 
\begin{equation}
	\langle k \rangle  \approx \gamma m \omega_y^2\,.
\end{equation}
Then gravity causes a  downward  displacement of the beam, which is given by
\begin{equation}
	\Delta y \approx  \frac{(2\gamma^2-1)\,|g_\oplus|}{\gamma^2\nu_y^2\,\omega_\text{rev}^2}\,,\label{eq:BeamShift1}
\end{equation}
where $\nu_y = \omega_y/\omega_\text{rev}$ is the vertical betatron tune and $\omega_\text{rev}$ is the beam orbital angular velocity.

Of particular interest is the all-electric frozen-spin ring for counter-rotating beams of protons with kinetic energy $T_p = \SI{233}{MeV}$. The often discussed lattice for such a ring with focusing by electric quadrupoles anticipates a circumference of \SI{500}{\meter} with $\omega_\text{rev} = \SI{2.26e6}{\per \second}$ and vertical betatron tune $\nu_y =0.45$\,\cite{srEDM,CPEDM}. The coherent vertical displacement of the beams induced by the Earth's gravity is therefore 
\begin{equation}
   \Delta y_{E} \approx \SI{13}{\pico \meter}\, .  
   \label{eq:yE}  
\end{equation}
In this scenario, both counter-rotating beams will have an identical gravitational displacement. As for the spin rotations of the protons, the impact of  the gravity-compensating electric field focusing in the vertical direction would correspond to a rotation of their magnetic dipole moments in a radial magnetic field of  opposite sign for clockwise and counter-clockwise propagating beams, respectively. This  would distinguish such a signal from a real EDM signal, which will be identical for both beams. However small the displacement may be, the false EDM effect due to the gravity-compensating electric field is even larger than the EDM effect expected  for  $d_p = \SI{e-29}{e.cm}$\,\cite{OrlovGravity,ObukhovSilenkoTeryaev,CPEDM}.  

A recent proposal of a hybrid electric ring with magnetic focusing\,\cite{OmarovHybrid} assumes a circumference of \SI{800}{\meter} with $\omega_\text{rev} = \SI{1.4e6}{\per  \second}$ and a vertical betatron tune  $\nu_y = 2.3$, giving a coherent gravitational vertical displacement of
\begin{equation}
\Delta y_B \approx \SI{1.3}{\pico \meter} \, ,
   \label{eq:yH}
\end{equation}
and an average beam splitting of \SI{2.6}{\pico \meter}. Note that the magnetic quadrupoles exert forces of opposite sign on counter-rotating beams, resulting in vertically non-identical orbits of the two beams. 

{\bf Classical mechanics of RF-driven beam oscillations} The treatment of the classical limit provides the necessary background and elucidates the subsequent transition to the deep quantum regime. Here we follow the discussion of Ref.\,\cite{Slim:2021wyk} and extend it to include the effects of intrabeam scattering and interaction with the residual gas. In the experiment described in Ref.\,\cite{Slim:2021wyk}, the beam was stroboscopically excited with a mismatched RF Wien filter once per turn. In this way, a vertical Lorentz force
\begin{equation}
F_y(n)= F_y \cos(n\, \omega_\text{WF} T) \,,\label{Force}
\end{equation} 
is exerted on the stored particle, where $n$ is the number of turns, $\omega_\text{WF}$ denotes the angular velocity of the RF in the Wien filter and $T = 2\pi/\omega_\text{rev}$ is the beam revolution period. When the Lorentz force vanishes, the beam performs idle vertical (and horizontal) betatron oscillations 
\begin{equation} 
	y(t) = y(0)\sqrt{\frac{\beta_y(t)}{\beta_y(0)}} \cos\left[ \psi_y(t) \right]\,,
	\label{beta-function}
\end{equation}
where $\beta_y(t)$ is the vertical betatron amplitude function. The betatron phase advance $\psi_y(t)$ satisfies $\psi_y(t+T) - \psi_y(t)=\omega_y T =2\pi \nu_y$. The  change at turn $n$ of the vertical velocity of the stored particle  accumulated during the time interval $\Delta t= \ell/v_z$ spent by the particle per turn inside the Wien filter of length $\ell$ is given by
\begin{equation}
	\Delta v_y(n) = \frac{ F_y (n) \Delta t }{\gamma m } = -\zeta \omega_y \cos(n\, \omega_\text{WF} T)  \,, 
	\label{eq:velKick}
\end{equation}
where again $\gamma$ and $m$ are the Lorentz factor and the mass of the particle, respectively. The change $\Delta y$ of the vertical position $y$ in the Wien filter can be neglected.
	
According to Eq.\,(\ref{beta-function}),  the stroboscopic signal of the betatron motion observed at any point in the ring follows the harmonic law as a function of $nT$ with angular velocity $\omega_y$, and we invoke the familiar description of the oscillatory motion in terms of the complex variable $z = y - iv_y/\omega_y$. The one-particle master equation for the buildup of RF-driven oscillations directly downstream of the Wien filter is
\begin{equation}
	     z(n) =z(n-1)\exp(i\omega_y T) -\frac{i}{\omega_y}\Delta v_y(n)\, . 
	     \label{MasterEqn}
\end{equation} 
Subject to the initial condition $z(0)=0$, it has   the generic solution 
\begin{equation}
		z(n)  =  -\frac{i}{\omega_y} \exp(i\omega_y nT)\sum_{k=1}^n \Delta v_y(k) \exp(-i\omega_y k T)\, ,
		\label{eq:GenericSolution}
\end{equation}
which yields for the stroboscopic force of Eq.\,(\ref{eq:velKick}) 
\begin{equation}
	\begin{split}
				z(n)  =  &\frac{i \zeta}{2}\cdot \frac{ \exp(i n\omega_y T) - \exp(i n \,\omega_\text{WF} T) }
				{\exp[i(\omega_y -\omega_\text{WF} )T]-1  }\\
		&+ \{ \omega_\text{WF} \to -\omega_\text{WF} \}\,.
		\label{eq:DrivenSolution}
	\end{split}
\end{equation}
A similar analytic result holds also for generic AC dipole-driven betatron oscillations, applied in a completely different context of machine diagnostics, described in Ref.\,\cite{MiyamotoACdipole}  (see also references therein). The amplitude of the RF Wien filter Fourier component of the beam oscillation, $y_\text{WF}(n) = - \xi_y\cos (n \, \omega_\text{WF} T)$, is given by
\begin{equation}
	\xi_y = \frac{\zeta}{2}\cdot \frac{\sin (2\pi \nu_y)}{\cos (2\pi \nu_\text{WF})  - \cos(2\pi \nu_y)}\, ,\label{eq:FourierAmplitude}
\end{equation}
where the WF tune $\nu_\text{WF}=\omega_\text{WF}/\omega_\text{rev}$. Note the resonance at $\nu_\text{WF} = \nu_y$.\footnote{In fact, the real part of $z(n)$,   as given in Eq.\,(\ref{eq:DrivenSolution}),  simply reads  $y(n) = - \zeta_y \bigl\{\cos(n\omega_\text{WF} T) - \cos(n\omega_yT) \bigr\} + \frac{\zeta}{2}\sin(n\omega_y T)$. Thus  on resonance,  we  obtain  $y^\text{res}(n) =- \frac{\zeta}{2}  (n-1) \sin(n\omega_yT)$.} The amplitude in Eq.\,(\ref{eq:FourierAmplitude}) is independent of the  betatron idle motion of individual particles and is shared by all particles in the bunch. It can be filtered out and measured by Fourier analysis of the BPM response with lock-in amplifiers. 
	 	
The Heisenberg uncertainty limit $Q$ for the betatron oscillation amplitude $\xi_y$ is related to the zero-point oscillator energy $\frac{1}{2} \hbar \omega_y$ through
\begin{equation}
	Q^2 = \frac{\hbar}{m \gamma \omega_ y}\, \label{eq:Heisenberg}	\,.
\end{equation}
Under the conditions of the experiment, this gives 
\begin{equation}
	Q=\frac{82}{\sqrt{\gamma \nu_y}}\,{\rm nm} = \SI{41}{\nano \meter}\, ,
	\label{eq:quantum-limit-1}
\end{equation}
while the smallest value of the  measured oscillation amplitude was $\xi_y^\text{min} = (1.08 \pm 0.52)\,\si{\micro \meter}$\,\cite{Slim:2021wyk}.

One can extend the above considerations to include the impact of  the interaction with the residual gas (RG). The scattering off the residual gas, followed by the vertical velocity kick $v_y^\text{RG}$ within the ring acceptance angle,  $\theta_\text{acc}$, is said to occur during the turn $n_\text{RG}$. According to Eq.\,(\ref{eq:GenericSolution}), the corresponding contribution to the betatron motion is
\begin{equation}
	z^\text{RG}(n)  =  -\frac{i}{\omega_y}\ v_y^\text{RG} 
	 \exp \left[i\omega_y (n-n_\text{RG})T) +\psi_y^\text{RG} \right]\, ,
	\label{eq:ResidualGas}
\end{equation}
where $\psi_y^\text{RG}$ is the betatron phase advance from the scattering point to the WF.	Evidently, scattering from the residual gas does not contribute at all to the RF-driven oscillations with angular velocity $\omega_\text{WF}$.  The same is true, of course, for intrabeam scattering. However, losses due to elastic scattering beyond the ring acceptance angle and the absorption due to electromagnetic and hadronic interactions must be taken into account by the obvious substitution in Eq.\,(\ref{eq:DrivenSolution}),		
\begin{equation}
		\omega_y \to \omega_y + \frac{i}{\tau}\,,
\end{equation}
where $\tau$ is the lifetime of the beam which amounts to $\approx  \SI{1500}{\second}$ (see \cite[Fig.\,14]{Slim:2021wyk}). The beam attenuation during one revolution is entirely negligible and the net effect on the oscillation amplitude is the well-known attenuation factor in Eq.\,(\ref{eq:FourierAmplitude}),
\begin{equation}
\xi_y \to \xi_y\exp\left(-\frac{nT}{\tau}\right)\, . \label{eq:Attenuation}
\end{equation}

\textbf{Quantum mechanics of RF Wien filter-driven oscillations} The above treatment of the proximity to the quantum limit by classical mechanics can be justified a posteriori, since the observed  amplitudes are much larger than the amplitude of the zero-point quantum  oscillations, given in Eq.\,(\ref{eq:quantum-limit-1}). Such a posteriori comparison of two amplitudes could have been  the other way around if the perturbation of the  beam had been much smaller than the zero-point quantum amplitude. Then one has to resort to the time-dependent Schr\"odinger equation
\begin{equation}
	i \frac{d}{dt}\Psi(t) = \left\{H_0 +V(t) \right\}\Psi(t) \, , \label{SchrEq}
\end{equation}
where $H_0$  is the time-independent potential of the harmonic oscillator, while the perturbative potential reads
\begin{equation}
	\begin{split}
	V(t)&=-F_y \cdot y \cdot\cos( \omega_\text{WF} t )\\
	& = -\frac{F_y}{\sqrt{2m\gamma\omega_y}}(a^\dagger +  a)\cos (\omega_\text{WF} t)\, , \label{Potential}
	\end{split}
\end{equation}
where $a^\dagger$ and $a$ are the harmonic oscillator creation and annihilation operators. This perturbation  stroboscopically acts for very short time intervals, $\Delta t \ll T$,  once per turn at time $t_n= nT$. 
	
Let $\Psi(-;n)$ be the wave function before, and $\Psi(+;n)$ directly behind the WF. The impact of the WF potential leads to  the following discontinuity of the wave function
\begin{equation}
	i\{ \Psi(+;n) -\Psi(-;n)\} = V(nT) \,\Delta t \, \Psi(-;n)\, , \label{Discontinuity}
\end{equation}
which gives 
\begin{equation}
	\begin{split}
	&\Psi(+;n) =\\ 
	&\quad\left\{1+i\frac{ F_y \Delta t}{\sqrt{2m\gamma \omega_y} }\cos(n\omega_\text{WF} T)
	 (a^\dagger +a)\right\} \Psi(-;n) \, . \label{Master}
	\end{split}
\end{equation}
The rest of the turn proceeds in the time-independent harmonic potential. The creation and annihilation operators change the energy of the state by $\pm \omega_y$. Taking this into account, we obtain the master equation
\begin{equation}
	\begin{split}
	&\Psi(-;n) = \\
	&\mbox{}\quad \left\{1 + i\frac{ F_y \Delta t}{\sqrt{2m\gamma \omega_y}} \cos(n\omega_\text{WF} T) 
	\left(a^\dagger e^{-i\omega_y T}  +a e^{i\omega_y T } \right) \right\}\\
	&\mbox{}\quad \times \Psi(-;n-1) e^{-i\omega_{\text{in}} T}\, . \label{HOevolution}
	\end{split}
\end{equation}
Here $\omega_\text{in}$ is the energy of the initial state $\Psi(+;0) = |\text{in}\rangle$.
	
A perturbative solution of this master equation proceeds as follows. The beam passes the Wien filter at times $t_k = kT$ with 
$k=1,2,\ldots, n$, and  the transition from the initial state $| \text{in}\rangle $ to the perturbation components $a^\dagger | \text{in}\rangle $ and $a | \text{in}\rangle$ can take place during any pass $k$.  Consequently, in the former case, the transition amplitude acquires   $\cos(k\omega_\text{WF} T)$ from the Wien filter potential and the phase factor $\exp(-i (n-k) \omega_y T)$  from the subsequent evolution in the confining potential. In the latter case the transition amplitude acquires   $\cos(k\omega_\text{WF} T)$ from the Wien filter potential and the phase factor $\exp(i (n-k) \omega_y T)$  from the subsequent evolution in the confining potential. Here  $\exp(i (n-k) \omega_y T)$ is the complex conjugate to $\exp(-i (n-k) \omega_y T)$, and upon summation of all transitions we obtain 
\begin{equation}
	\begin{split}
	|\Psi(+;n)\rangle & =\left\{ 1  + i \frac{F_y \Delta t}{\sqrt{2\gamma m \omega_y} }
	\left(w(n) a^\dagger  +w^*(n) a \right) \right\} |\text{in}\rangle\, , 
	\label{GenericMasterEq}
	\end{split} 
\end{equation}
where
\begin{equation}
	\begin{split}
	w(n)&= \sum_{k=1}^{n} \cos (k\omega_\text{WF} T)\exp\{-i(n-k)\omega_y T\}\\
	&=\frac{1}{2}\cdot\Bigl[  
	\frac{\exp(-i n\omega_y T)- \exp(-i n \omega_\text{WF} T)}
	{\exp (-i(\omega_y-\omega_\text{WF}) T)-1}\\
	&\qquad+\{\omega_\text{WF} \to -\omega_\text{WF}  \}\Bigr]\, . \label{SummedAmplitude}
	\end{split} 
\end{equation}
	
Now we are in the position to evaluate the driven oscillation amplitude. To linear order in $\Delta t$ we get 
\begin{equation}
	\begin{split}
	y(n) &= \frac{1}{\sqrt{2\gamma m \omega_y}} \left\langle \Psi^*(+;n) \left| (a^\dagger  + a) \right| \Psi(+;n) \right\rangle \\ 
	&=  i\frac{F_y \Delta t}{2\gamma m \omega_y}
	\left\langle \text{in} \left|\left[ \left(a^\dagger  +a\right),\left( w(n) a^\dagger  + w^*(n) a\right) \right] \right |\text{in} \right\rangle\\
	&=  -i\frac{F_y \Delta t}{2\gamma m \omega_y} \left(w^*(n)-w(n)\right)
	\left \langle \text{in}  \left |[a,a^\dagger] \right | \text{in} \right\rangle\\
	&=  -i\frac{F_y \Delta t}{2\gamma m \omega_y}\left (w^*(n)-w(n) \right)\,.
	\label{eq:ynQM}
	\end{split}
\end{equation}
This expression is exactly the same as for  $y(n) = {\rm Re}\, z(n)$ from Eq.\,(\ref{eq:DrivenSolution}),
with $\zeta$ replaced by  the ratio $ \frac{-F_y\Delta t}{\gamma m \omega_y}$, {\it cf.} 
Eq.\,(\ref{eq:velKick}).

The appearance of the commutator $[a,a^\dagger]=1$ in Eq.\,(\ref{eq:ynQM}) is crucial. It makes the driven coherent oscillation amplitude universal for all particles in the beam, regardless of their individual quantum state. This finding is tantamount  to the independence of the classical driven oscillation amplitude from the amplitude and phase of the inherent betatron motion\,\cite{Slim:2021wyk}. The amplitude of the RF Wien filter-driven oscillations of the bunch is identical to the amplitude of each individual particle in the bunch. Moreover, quantum mechanics and classical mechanics give absolutely identical results for the amplitude of driven oscillations. One may consider this as an exemplary case of Ehrenfest's theorem.
	
{\bf Summary and Conclusions}
We have presented a quantum mechanical description of the excitation of coherent betatron oscillations by 
radiofrequency electromagnetic fields. Remarkably, one and the same formula [Eq.\,(\ref{eq:FourierAmplitude})] covers the whole range of amplitudes from  large classical ones  to well below the one-particle quantum limit. Neither scattering from the residual gas nor intrabeam scattering contribute to these coherent betatron
oscillations.
	
We conclude that in principle the amplitude of coherent oscillations of the center of mass of a particle bunch in a storage ring can be measured with an accuracy of better than one picometer within the framework of the Heisenberg uncertainty principle. Our analysis may be applied to other problems involving  pulsed excitation of quantum oscillators. 

\begin{acknowledgments}
This work has been performed in the framework of the JEDI collaboration, and is financially supported by an ERC Advanced-Grant (srEDM \# 694390) of the European Union and by the Russian Fund for Basic Research (Grant No. 18-02-40092 MEGA). 
\end{acknowledgments} 	

\bibliography{QuantumBibUpdate}
\end{document}